\documentclass{emulateapj}
\usepackage{natbib}

\begin{document}
\title{Discovery of VHE $\gamma$-ray emission from the SNR G54.1+0.3}

\author{
V.~A.~Acciari\altaffilmark{1},
E.~Aliu\altaffilmark{2, $\amalg$, *},
T.~Arlen\altaffilmark{3},
T.~Aune\altaffilmark{4},
M.~Bautista\altaffilmark{5},
M.~Beilicke\altaffilmark{6},
W.~Benbow\altaffilmark{1},
D.~Boltuch\altaffilmark{2},
S.~M.~Bradbury\altaffilmark{7},
J.~H.~Buckley\altaffilmark{6},
V.~Bugaev\altaffilmark{6},
Y.~Butt\altaffilmark{8},
K.~Byrum\altaffilmark{9},
A.~Cesarini\altaffilmark{10},
L.~Ciupik\altaffilmark{11},
W.~Cui\altaffilmark{12},
R.~Dickherber\altaffilmark{6},
C.~Duke\altaffilmark{13},
J.~P.~Finley\altaffilmark{12},
G.~Finnegan\altaffilmark{14},
L.~Fortson\altaffilmark{11},
A.~Furniss\altaffilmark{4},
N.~Galante\altaffilmark{1},
D.~Gall\altaffilmark{12},
G.~H.~Gillanders\altaffilmark{10},
S.~Godambe\altaffilmark{14},
E.~V.~Gotthelf\altaffilmark{15},
J.~Grube\altaffilmark{16,11},
R.~Guenette\altaffilmark{5},
G.~Gyuk\altaffilmark{11},
D.~Hanna\altaffilmark{5},
J.~Holder\altaffilmark{2},
C.~M.~Hui\altaffilmark{14},
T.~B.~Humensky\altaffilmark{17},
A.~Imran\altaffilmark{18},
P.~Kaaret\altaffilmark{19},
N.~Karlsson\altaffilmark{11},
M.~Kertzman\altaffilmark{20},
D.~Kieda\altaffilmark{14},
A.~Konopelko\altaffilmark{21}
H.~Krawczynski\altaffilmark{6},
F.~Krennrich\altaffilmark{18},
M.~J.~Lang\altaffilmark{10},
S.~LeBohec\altaffilmark{14},
G.~Maier\altaffilmark{5, $\diamond$},
S.~McArthur\altaffilmark{6},
A.~McCann\altaffilmark{5},
M.~McCutcheon\altaffilmark{5},
P.~Moriarty\altaffilmark{22},
R.~Muhkerjee\altaffilmark{23}
R.~A.~Ong\altaffilmark{3},
A.~N.~Otte\altaffilmark{4},
D.~Pandel\altaffilmark{19},
J.~S.~Perkins\altaffilmark{1},
M.~Pohl\altaffilmark{18, $\diamond$,$\aleph$},
J.~Quinn\altaffilmark{16},
K.~Ragan\altaffilmark{5},
L.~C.~Reyes\altaffilmark{24},
P.~T.~Reynolds\altaffilmark{25},
E.~Roache\altaffilmark{1},
H.~J.~Rose\altaffilmark{7},
M.~Schroedter\altaffilmark{18},
G.~H.~Sembroski\altaffilmark{12},
G.~Demet~Senturk\altaffilmark{15},
P.~Slane\altaffilmark{8},
A.~W.~Smith\altaffilmark{9},
D.~Steele\altaffilmark{11, $\S$},
S.~P.~Swordy\altaffilmark{17},
G.~T\v{e}si\'{c}\altaffilmark{5},
M.~Theiling\altaffilmark{1},
S.~Thibadeau\altaffilmark{6},
V.~V.~Vassiliev\altaffilmark{3},
S.~Vincent\altaffilmark{14},
S.~P.~Wakely\altaffilmark{17, *},
J.~E.~Ward\altaffilmark{16},
T.~C.~Weekes\altaffilmark{1},
A.~Weinstein\altaffilmark{3},
T.~Weisgarber\altaffilmark{17},
D.~A.~Williams\altaffilmark{4},
S.~Wissel\altaffilmark{17},
M.~Wood\altaffilmark{3},
B.~Zitzer\altaffilmark{12}
}

\altaffiltext{1}{Fred Lawrence Whipple Observatory, Harvard-Smithsonian Center for Astrophysics, Amado, AZ 85645, USA}
\altaffiltext{2}{Department of Physics and Astronomy and the Bartol Research Institute, University of Delaware, Newark, DE 19716, USA}
\altaffiltext{3}{Department of Physics and Astronomy, University of California, Los Angeles, CA 90095, USA}
\altaffiltext{4}{Santa Cruz Institute for Particle Physics and Department of Physics, University of California, Santa Cruz, CA 95064, USA}
\altaffiltext{5}{Physics Department, McGill University, Montreal, QC H3A 2T8, Canada}
\altaffiltext{6}{Department of Physics, Washington University, St. Louis, MO 63130, USA}
\altaffiltext{7}{School of Physics and Astronomy, University of Leeds, Leeds, LS2 9JT, UK}
\altaffiltext{8}{Harvard-Smithsonian Center for Astrophysics, 60 Garden Street, Cambridge, MA 02138, USA}
\altaffiltext{9}{Argonne National Laboratory, 9700 S. Cass Avenue, Argonne, IL 60439, USA}
\altaffiltext{10}{School of Physics, National University of Ireland Galway, University Road, Galway, Ireland}
\altaffiltext{11}{Astronomy Department, Adler Planetarium and Astronomy Museum, Chicago, IL 60605, USA}
\altaffiltext{12}{Department of Physics, Purdue University, West Lafayette, IN 47907, USA }
\altaffiltext{13}{Department of Physics, Grinnell College, Grinnell, IA 50112-1690, USA}
\altaffiltext{14}{Department of Physics and Astronomy, University of Utah, Salt Lake City, UT 84112, USA}
\altaffiltext{15}{Columbia Astrophysics Laboratory, Columbia University, New York, NY 10027, USA}
\altaffiltext{16}{School of Physics, University College Dublin, Belfield, Dublin 4, Ireland}
\altaffiltext{17}{Enrico Fermi Institute, University of Chicago, Chicago, IL 60637, USA}
\altaffiltext{18}{Department of Physics and Astronomy, Iowa State University, Ames, IA 50011, USA}
\altaffiltext{19}{Department of Physics and Astronomy, University of Iowa, Van Allen Hall, Iowa City, IA 52242, USA}
\altaffiltext{20}{Department of Physics and Astronomy, DePauw University, Greencastle, IN 46135-0037, USA}
\altaffiltext{21}{Department of Physics, Pittsburg State University, 1701 South Broadway, Pittsburg, KS 66762, USA}
\altaffiltext{22}{Department of Life and Physical Sciences, Galway-Mayo Institute of Technology, Dublin Road, Galway, Ireland}
\altaffiltext{23}{Department of Physics and Astronomy, Barnard College, Columbia University, NY 10027, USA}
\altaffiltext{24}{Kavli Institute for Cosmological Physics, University of Chicago, Chicago, IL 60637, USA}
\altaffiltext{25}{Department of Applied Physics and Instrumentation, Cork Institute of Technology, Bishopstown, Cork, Ireland}
\altaffiltext{$\amalg$}{ Now at Department of Physics and Astronomy, Barnard College, Columbia University, NY 10027, USA}
\altaffiltext{$\diamond$}{ Now at DESY, Platanenallee 6, 15738 Zeuthen, Germany}
\altaffiltext{$\aleph$}{ Now at Institut f\"ur Physik und Astronomie, Universit\"at Potsdam, 14476 Potsdam-Golm, Germany}
\altaffiltext{$\S$}{ Now at Los Alamos National Laboratory, MS H803, Los Alamos, NM 87545, USA}
\altaffiltext{*}{Address correspondance to E.~Aliu or S.~P.~Wakely \\
E-mail: ealiu@astro.columbia.edu and wakely@uchicago.edu}

%%%%%%%%%%%%%%%%%%%%%%%%%%%%%%%%%%%%%%%%%%%%%%%%%%%%%%%%%%%%%%%%%%%%%%%%%%%%%%%
%%%%%%%%%%%%%%%%%%%%%%%%%%%%%%%%%%%%%%%%%%%%%%%%%%%%%%%%%%%%%%%%%%%%%%%%%%%%%%%
\begin{abstract}

We report the discovery of very high energy gamma-ray emission from the direction of the SNR G54.1+0.3 using the VERITAS ground-based gamma-ray observatory.~The TeV signal has an overall significance of 6.8$\sigma$ and appears point-like given the %5$^{arcminute}$
resolution of the instrument.~The integral flux above 1 TeV is 2.5\% of the Crab Nebula flux and significant emission is measured between 250 GeV and 4 TeV, well described by a power-law energy spectrum dN/dE $\sim$ E$^{-\Gamma}$ with a photon index $\Gamma= 2.39\pm0.23_{stat}\pm0.30_{sys}$.~We find no evidence of time variability among observations spanning almost two years.~Based on the location, the morphology, the measured spectrum, the lack of variability and a comparison with similar systems previously detected in the TeV band, the most likely counterpart of this new VHE gamma-ray source is the PWN in the SNR G54.1+0.3.~The measured X-ray to VHE gamma-ray luminosity ratio is the lowest among all the nebulae supposedly driven by young rotation-powered pulsars, which could indicate a particle-dominated PWN.   

\end{abstract}


\section{Introduction}

In recent years, new insights into the most extreme environments in the Galaxy have been obtained through the very high energy~(VHE, $>$100 GeV)~gamma-ray band~\citep{aharonian05a}.~The majority of the $>$50 Galactic TeV sources found thus far are associated with supernova remnants (SNRs) or pulsar wind nebulae (PWNe)\footnote{TeVCat, an online catalog for TeV Astronomy, http://tevcat.uchicago.edu/}.~The detection of these stellar remnants in VHE gamma rays constitutes a probe of non-thermal astrophysical processes.~The acceleration of  electrons or nuclei, above 1 TeV,  in these sources leads to the generation of gamma rays.~Relativistic electrons can produce gamma rays by non-thermal bremsstrahlung, and by inverse-Compton (IC) scattering on ambient microwave, IR, or optical photons, whereas protons and atomic nuclei can generate gamma rays via the decay of neutral pions produced in hadronic interactions with ambient material.~A confirmation of the latter process associated with the gamma rays from SNRs is seen as a probe of nucleonic cosmic ray production sites in the Galaxy.
\par
The SNR G54.1+0.3 has been predicted to emit gamma rays detectable with the current generation of TeV instruments after the discovery of a rotation-powered pulsar in its centre, a source of high energy electrons~(e.g.~\cite{bednarek05}).~This pulsar, PSR J1930+1852~(P=136 ms), with a derived spindown power of $\dot{E}$ = 1.2$\times10^{37}$ erg/s and a spindown age of $\tau_c$$\sim$2900 years~\citep{camilo02}, is among the youngest and most energetic pulsars known.~The PWN powered by this pulsar was first recognized as such from high resolution radio observations~\citep{reich85}.~The PWN exhibits a filled-center morphology~($2'\times1.5'$), a flat radio spectrum~($\alpha\approx0.13$) which is fairly constant over the entire source extent, and a high level of polarization~\citep{velusamy88}.~The discovery of an X-ray jet,~consistent with the radio extension,~and the measurement of its non-thermal spectrum confirmed the PWN nature of G54.1+0.3~\citep{lu01}.~More detailed X-ray observations by~\cite{lu02} resolved the inner structure of the nebula as a central point-like source with a surrounding ring and bipolar elongations, comparable to the Crab Nebula.~The same authors suggested a distance of $\sim$5 kpc, based on the X-ray absorption.~The most recent estimate is 6.2$^{+1.0}_{-0.6}$ kpc, based on an apparent morphological association between the PWN and a nearby molecular CO cloud~\citep{leahy08}.
\par
Similar to the Crab Nebula, G54.1+0.3 lacks a strong surrounding circumstellar interaction between the SNR forward shock and the progenitor star wind~\citep{chevalier05}, in which particles could be accelerated and produce an additional non-thermal emission contribution.~Recently, \cite{leahy08} found a shell of 30' diameter surrounding G54.1+0.3, but the authors concluded that these are unlikely to be associated based on the different measured distances.~After that,~\cite{koo08} identified a partial shell-like infrared structure surrounding much of the X-ray/radio core in G54.1+0.3, which they proposed to be a star-forming region triggered by late-phase winds from the SNR progenitor star.~Based on high-resolution IR spectra, however,~\cite{temim09} suggested that the shell is composed of freshly-formed ejecta dust that appears to be heated by the expanding PWN.
 \par
G54.1+0.3 has not been previously detected at $\gamma$-ray energies.~The source lies 1.4$^{\circ}$ from the EGRET source  3EG J1928+1733~\citep{hartman99}.~This is slightly outside the 99\% confidence-level error box.~The older~($\tau_{c}$$\sim$8.2$\times10^{4}$ years)~and less-energetic ($\dot{E}$ = 1.6$\times10^{36}$ ergs/s) pulsar PSR J1928+1746 is proposed as a more likely counterpart to the unidentified 3EG source~\citep{cordes06}.~This is strengthened with the much better localization of the gamma-ray source 1FGL J1929.0+1741c,~associated with 3EG J1928+1733 in the first Fermi catalogue~\citep{FermiCatalogue}.~This other pulsar is 1.2$^{\circ}$ away from PSR J1930+1852.~In the VHE $\gamma$-ray domain, the only published observation is an upper limit from the HEGRA collaboration at $20\%$ of the Crab Nebula flux above 600 GeV~\citep{aharonian02}.~In this Letter we report the discovery by VERITAS of VHE gamma-ray emission from the direction of the very young system G54.1+0.3/PSR J1930+1852.~The instrument and observations are described in \S~2,~while the analysis performed on these data is detailed in \S~3.~Results are presented in \S~4 and interpretation and conclusion in \S~5. 

\section{Observations}

VERITAS is an array of four imaging atmospheric Cherenkov telescopes located at the Fred Lawrence Whipple Observatory in southern Arizona, USA \citep{holder06, holder08}.~Each telescope has a mirror area of 110 m$^2$ and is equipped with a 499-pixel camera of 3.5$^{\circ}$ diameter field-of- view.~The system, completed in the fall of 2007,~is run in a coincident mode requiring at least two of the four telescopes to trigger in each event.~This design allows for the observations of astrophysical objects in the energy range from 100 GeV to above 30 TeV.~VERITAS has an angular resolution (68\% containment) of 0.1$^{\circ}$ diameter per event and a 5$\sigma$ point source sensitivity of 1\% of the steady Crab Nebula flux above 300 GeV in less than 50 hours of observation at a 20$^{\circ}$ zenith angle. 
\par
The first pointed observations of the G54.1+0.3/PSR J1930+1852 system with VERITAS took place during a survey of northern-hemisphere Galactic pulsars between September 2007 and October 2008.~The live time of this first data set, after data-quality selection based on weather and hardware status, amounts to 14.5 hr (some of which was taken in partial moonlight).~Analysis of these data revealed evidence of a VHE $\gamma$-ray signal at the position of PSR J1930+1852 at the 4$\sigma$ level after accounting for all analysis trials.~This evidence triggered further observations between April 2009 and June 2009, resulting in 22.2 additional hours after quality selection.~Of these, the most recent 15.2 hours of data were taken using only three telescopes, as telescope T1 was in the process of being relocated to optimize the instrument sensitivity~\citep{perkins09}.~The zenith angle range of observation of the combined data sets is 13$^{\circ}$ to 35$^{\circ}$, and the total live time is 36.6 hr. 

%%%%%%% TABLE 2.  Analysis Results %%%%%
%%%%%%%%%%%%%%%%%%%%%%%%
\begin{deluxetable*}{cccccccc}[t]
\tabletypesize{\scriptsize}
\tablecaption{Analysis results at the high Edot pulsar locations in the FOV of the PWN G54.1+0.3\label{table2}}
\tablecolumns{8}
\tablewidth{7in}
\tablehead{
\colhead{PSR name} & \colhead{Offset} &\colhead{On} & \colhead{Off} & \colhead{$\alpha$\tablenotemark{a}} & \colhead{Excess} &
\colhead{LiMa} & \colhead{Flux $>$1 TeV} \\
\colhead{ } & \colhead{[$^{\circ}$]}& \colhead{Events} & \colhead{Events} & \colhead{ } & \colhead{Events} &
\colhead{Significance} & \colhead{[10$^{-13}cm^{-2}s^{-1}$]} 
}
\startdata
J1930+1852 &  0.5 & 231 & 720 & 0.18 &  101 & 7.0$\sigma$ & 5.3 $\pm$ 1.3   \\ 
J1928+1746 &  0.7  to 1.7 & 108 & 509 & 0.19 & 13 & 1.2 $\sigma$ & $<2.6$ \\ 
\enddata
\tablecomments{The analysis used for the results presented in this table corresponds to an image integrated charge $\sim$225 photoelectrons and $\theta^2<0.015$ deg$^2$} 
\tablenotetext{a}{ Normalization factor for the different acceptance of the background and source regions in the ring background model }
\end{deluxetable*}

\section{Analysis}

After calibration, the standard VERITAS event reconstruction scheme is applied to the data: tail-cuts image-cleaning, second-moment parameterization of these images~\citep{hillas85} and stereoscopic direction determination based on the intersection of image axes (see e.g. \cite{krawczynski06} for details).~An initial selection of the events is performed to reject those in which the primary parameters, energy and direction, cannot be well-reconstructed.~The events surviving this quality selection are those for which at least two telescope images have more than 4 pixels included in the image with a total integrated charge in each image of more than 80 photoelectrons, and with less than 20\% of the total charge in the outermost ring of camera pixels.~This selection removes images with insufficient Cherenkov light for accurate parameterization, or which are truncated by the camera boundary.~In addition, events for which only the telescopes T1 and T4 have surviving images are excluded, as their separation of only 35 m introduces an irreducible high background rate, due to local muons, degrading the instrument sensitivity~\citep{maier07}.
\par
A further selection of events is performed to suppress the otherwise overwhelming hadronic cosmic-ray background.~This is achieved by comparing the shapes of the event images in each telescope with the expected shapes of simulated gamma-ray showers.~These {\it mean-reduced-scaled width} (MRSW) and {\it mean-reduced-scaled length} (MRSL) cuts (see definition in \citet{krawczynski06}), as well as an additional cut on the arrival direction of the incoming gamma-ray with respect to the source position~($\theta^2$) reject more than 99.9\% of the hadronic cosmic-ray background while keeping 45\% of the gamma-rays.~Different sets of these cuts, optimized for different source properties, are applied to the G54.1+0.3 data.~Two {\it a priori} test positions are defined as the locations of the high spindown power pulsars: J1930+1852 ($19^h30^{m}30.13^{s},+18^{\circ}52^{\prime}14.1^{\prime\prime}$) and J1928+1746 ($19^h28^{m}42.48^{s},+17^{\circ}46^{\prime}27^{\prime\prime}$).~In the search for a VHE $\gamma$-ray signal from a point source with a Crab-like spectrum the following cuts were employed: -1.2 $<$ MSCW/L $<$ 0.5 and $\theta^2<0.015$ deg$^2$, resulting in an analysis threshold of 250 GeV.~Since the emission region may be extended, as is the case of most of the galactic PWN detected at VHE $\gamma$-rays~\citep{aharonian05a}, a larger cut in the arrival direction $\theta^2<0.055$ deg$^2$ is also considered. In addition, a rather tight cut on the integrated charge of $\sim$225 photoelectrons is applied, to maximize the signal-to-noise ratio for a weak source with a hard spectrum.~The latter cut results in an analysis threshold of 500 GeV and in superior angular resolution.~For the spectral analysis, this cut is 80 photoelectrons providing maximal coverage in energy. 

\begin{figure}[tb]
\includegraphics[width=8.3cm,angle=90]{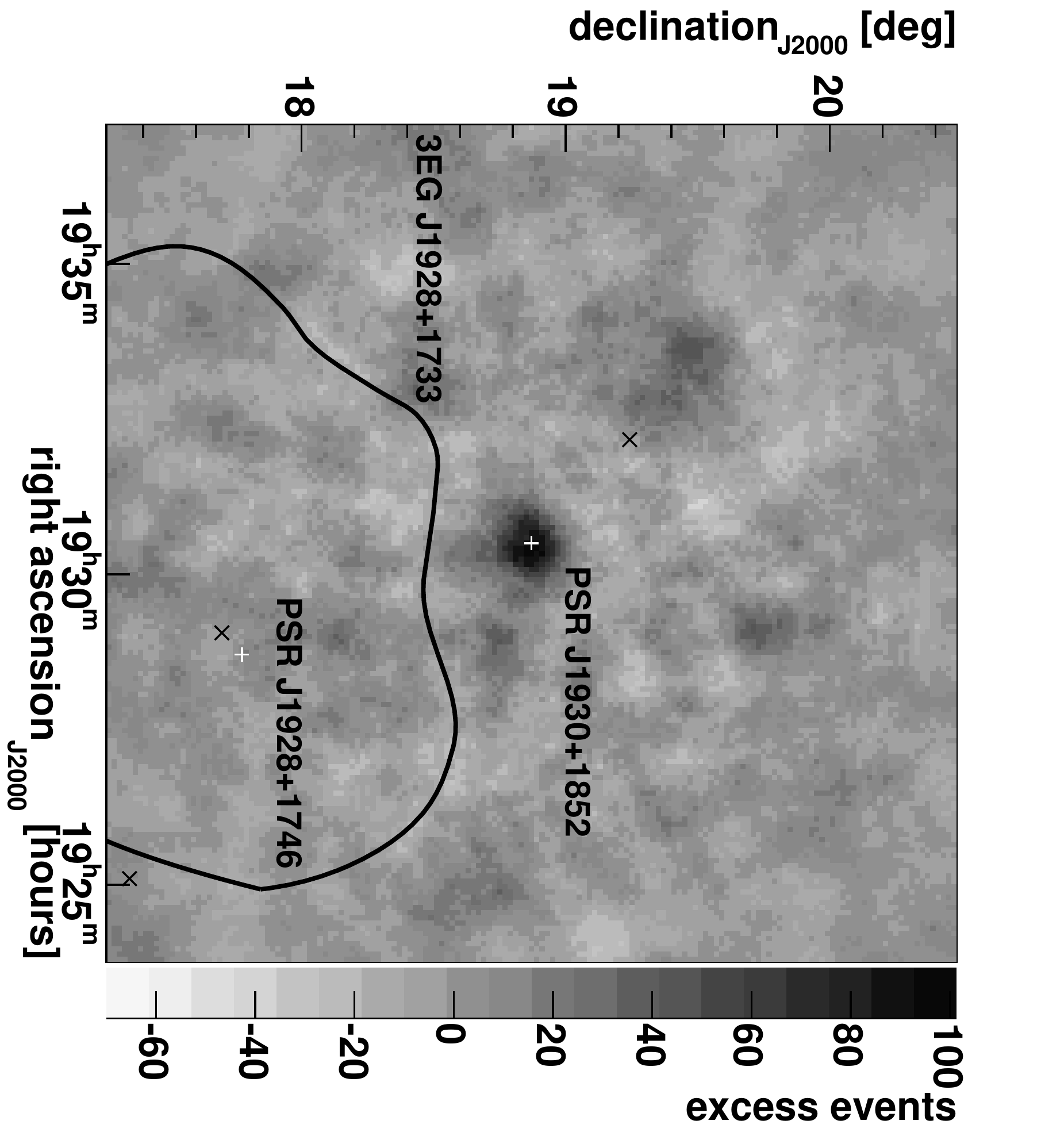}
\caption{\label{fig:map} Smoothed excess map from G54.1+0.3, as measured by VERITAS.~Only events with image sizes above 225 photoelectrons are used.~The color scale indicates the number of excess gamma-ray events.~The high $\dot{E}$ pulsars present in this field-of-view, J1930+1852 and J1928+1746, are marked with white plus signs.~The contour of the 99\% confidence level error box of the unidentified EGRET source 3EG J1928+1746 is overlaid in black. The gamma-ray sources in the first Fermi catalogue are marked with black crosses~\label{fig1}.}
\end{figure}

\par
For the background estimation, we employ two different methods, as described in~\citet{berge07}.~The background estimate for each position in the two-dimensional sky map is taken from a ring of mean radius 0.5$^{\circ}$ and an area six times that of the on-source region.~For the spectral analysis, the background is taken from 7 positions in the field-of-view with the same offset from the pointing direction as the source regions. This eliminates the need for corrections concerning the radial dependence of the background acceptance.~The on-source and background region counts, together with a normalization factor for the different acceptance of these regions, are used to derive the statistical significance of any excess following the likelihood method of \citet{lima83}.

\section{Results}

Figure~\ref{fig1} shows a sky map of the excess counts for the entire data set in the region around G54.1+0.3/PSR J1930+1852 system, derived using the point source analysis described above.~A source of VHE gamma rays is clearly visible and is coincident with the position of the X-ray and radio PWN.~An excess of 101 events is observed, corresponding to a pre-trials statistical significance of 7.0$\sigma$ at the location of PSR J1930+1852, as shown in Table~\ref{table2}.~This significance is the highest obtained at the pulsar position for the four different search criteria attempted, and this is taken into account in assessing the true chance probability of the observed signal of 6.8$\sigma$.~As discussed above, the earlier data set showed evidence for a VHE $\gamma$-ray signal at 4.3$\sigma$.~This evidence is confirmed by the presence of a 5.5$\sigma$ detection at the same position in 2009 data set alone.~This increase is as expected for a steady source with increased exposure time and a slightly reduced sensitivity of the array with only three telescopes. 
\par
The unsmoothed map of excess events, binned in 0.05$^{\circ}\times$0.05$^{\circ}$, is fit by a two-dimensional Gaussian in order to study the morphology of the observed emission.~The gamma-ray excess is well fit by the point-spread function of the instrument at the corresponding analysis threshold of 500 GeV ($\sim$0.11$^{\circ}$), which results in a detection compatible with a point-like source for VERITAS.~The best fit of the Gaussian centroid is $19^h30^{m}32^{s}\pm25^{s},+18^{\circ}52'12"\pm20"$ (J2000) and hence we assign the name VER J1930+188.~Given the systematic uncertainty of 0.02$^{\circ}$ in this measurement, the centroid is consistent with the pulsar position (at a distance of 0.007$^{\circ}$) and the 0.03$^{\circ}$ extent of the PWN G54.1+0.3 around the pulsar. 
\par
\begin{figure} [t]
\includegraphics[width=8.3cm,angle=90]{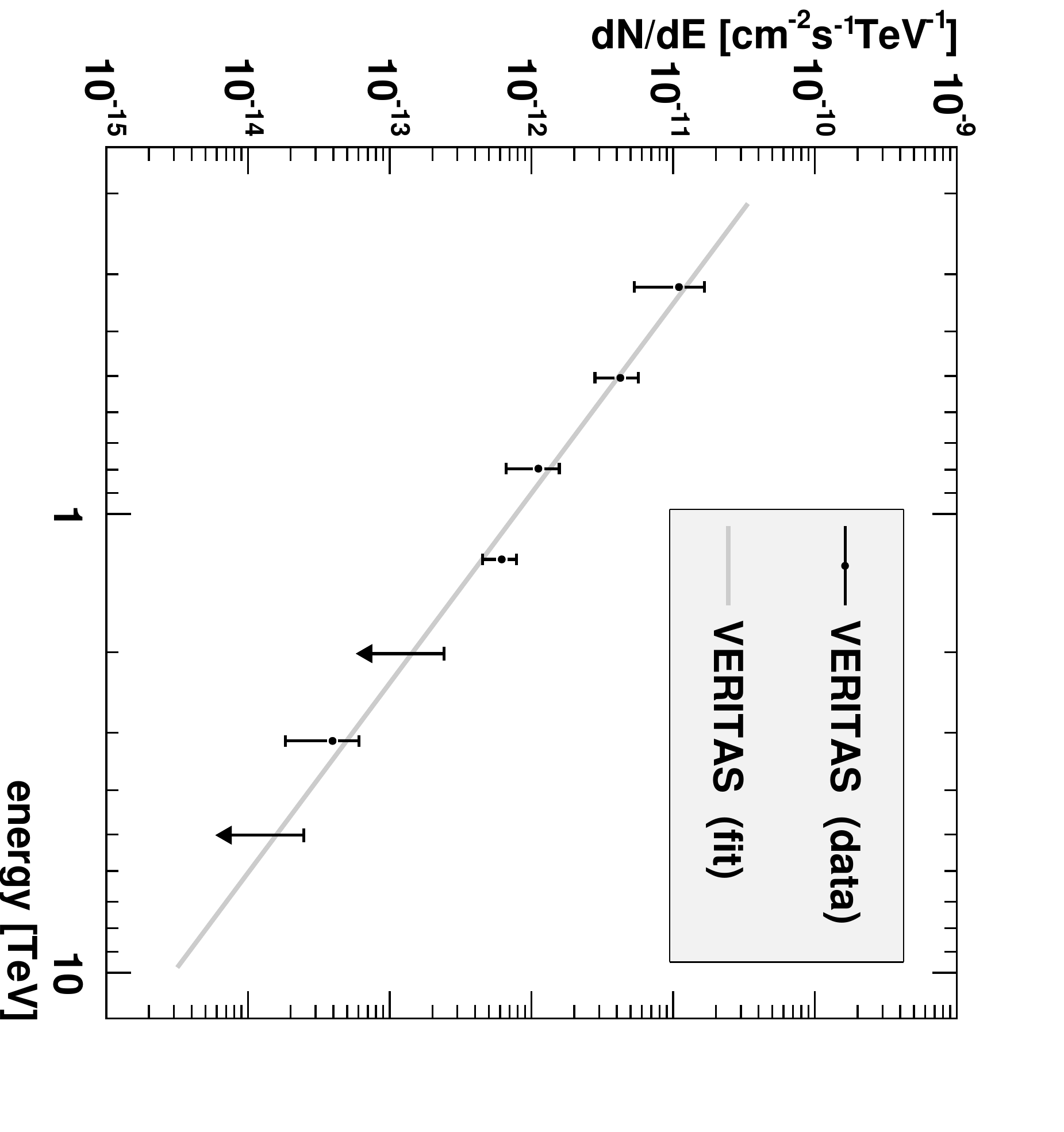}
\caption{Reconstructed VHE $\gamma$-ray spectrum of the new source VER J1930+188. The 99\% confidence level are determined with the method of \citet{feldman98} and are not used in the fit.~The parameters of the fitted power law can be found in the text.\label{fig2}}
\end{figure}

The total $\gamma$-ray excess above 250 GeV, i.e. that within a circle of 0.15$^{\circ}$ around the pulsar position, is  214$\pm$43 events.~Figure~\ref{fig2} shows the reconstructed gamma-ray spectrum from these events.~The data are consistent with a power law in energy that extends from 250 GeV up to 4 TeV with a photon index of $\Gamma = 2.39\pm0.23_{stat}\pm0.30_{sys}$ and a differential flux at 1 TeV of $(7.5\pm1.2_{stat}\pm1.5_{sys} )\times10^{-13}$ TeV$^{-1}$ cm$^{-2}$ s$^{-1}$.~The corresponding flux above 1 TeV is $~2.5\%$ of the Crab Nebula flux.

\par
No other significant TeV source is found in the maps, including at the second {\it a priori} test position, the location of PSR J1928+1746, for any of the searches performed.~The significance at this pulsar position is 1.2$\sigma$ and the flux upper limit above 1 TeV at the 99\% confidence level~\citep{feldman98} assuming a power law  source spectrum  with $\Gamma=2.5$ is F $<$~$2.6\times10^{-13}$ cm$^{-2}$ s$^{-1}$ (about 1.4\% of the flux of the Crab Nebula; see Table~\ref{table2}). We have used the point source analysis for these calculations, as there is no known X-ray or radio PWN or SNR counterpart associated with this pulsar which can be used to constrain the size of the TeV emission.  

%%%%%%% TABLE 3.  Comparison %%%%%
%%%%%%%%%%%%%%%%%%%%%%%%
\begin{deluxetable*}{ccccccccc}[t]
%\tabletypesize{\scriptsize}
\tablecaption{Pulsar and nebular properties of compact PWN detected at TeV \label{table3}}
\tablecolumns{7}
\tablewidth{4.5in}
\tablehead{
\colhead{Name} & \colhead{$\dot{E}$} & $\tau_c$ & \colhead{Distance} & \colhead{L$_\gamma$\tablenotemark{a}} &  \colhead{L$_X$\tablenotemark{b}}  &\colhead{ $\eta_\gamma$\tablenotemark{c}} & \colhead{ $\eta_X$\tablenotemark{d} }& \colhead{L$_X$/L$_\gamma$} \\
\colhead{} & \colhead{($10^{36}$ erg/s)} & \colhead{kyr} & \colhead{(kpc)} & \colhead{($10^{34}$ erg/s)} &  \colhead{($10^{34}$ erg/s)}  & \colhead{(\% )} &\colhead{(\% )} & \colhead{} \\
}
\startdata 
Crab Nebula  &   460  &   1.2    &   2      &   8.6    &   1900   &  0.02  &   4    &   220 \\
G0.9+0.1         &   43    &   5.3    &   8.5   &   4.4    &   36    &  0.1    &   0.8 &    8     \\
G21.5-0.9        &   33    &   4.85  &  4.7   &   0.72  &   22    &   0.02 &   0.7 &    30   \\
%MSH 15-52     &   18    &   1.55  &   5.2   &  11      &   0.37  &  0.6    &   0.2 &   0.3  \\
G54.1+0.3       &   12    &   2.9    &   6.2   &  2.1     &   2.9  &  0.17  &   0.2 &   1.4  \\
Kes75              &    8.1  &   0.7    &   7.5    &  2.5   &   24  &   0.3       &   2.9  &   10  \\
\enddata
\tablenotetext{a}{ Luminosity in the 0.2-10 TeV band }
\tablenotetext{b}{ Luminosity of the PWN in the 0.5-8 keV band, taken from \cite{kargalstev08} }
\tablenotetext{c}{ Efficiency to convert the spin-down energy to VHE gamma-ray emission  : $\eta_\gamma$$\equiv$L$_\gamma$/$\dot{E}$ }
\tablenotetext{d}{ Efficiency to convert the spin-down energy to X-ray synchrotron : $\eta_X$$\equiv$L$_X$/$\dot{E}$ }
\end{deluxetable*}
%%%%%%%%%%%%%%%%%%%%%%%%

\section{Discussion and Conclusion}
VERITAS has detected VHE gamma-ray emission from  the direction of the SNR G54.1+0.3, providing a new input to study the non-thermal particle population in this object.~This new TeV source, VER J1930+188,~appears point-like within the instrument resolution and does not show a notable offset from the pulsar PSR J1930+1852, placing it among the more compact TeV sources found to be associated with an SNR.~The emission is co-located with the X-ray and radio PWN in the SNR G54.1+0.3,~and also with a molecular CO cloud in apparent morphological association with the PWN and a partial shell of IR ejecta around the PWN.

\par
Assuming the gamma-ray production in VER J1930+188 is related to the pulsar, the compactness of the TeV source and lack of significant offset between the pulsar and its X-ray PWN and the TeV source can be attributed to its relative youth.~The spin-down age of PSR J1930+1852 is 2900 years, which places it as the latest addition to an emergent  class of TeV sources associated with very young pulsars:  PSR B0531+21/Crab Nebula~\citep{weekes89}, PSR J1747-2809/HESS J1747-281 in G0.9+0.1~\citep{camilo09,aharonian05b}, PSR J1845-0258/HESS J1845-029 in SNR Kes 75, PSR J1833-1036/HESS J1833-105 in G21.5-0.9~\citep{djannati07} and PSR J1813-1749/HESS J1813-178 in G12.82-0.02~\citep{gotthelf09}.~Those PWNe are good candidates for being in the early stage of evolution, where the wind nebula is interacting with the freely-expanding supernova ejecta.~This population stands in contrast to an apparently more common variety of energetic-pulsar-associated TeV sources, a group which contains, as notable examples, PSR B1823-1313/HESS J1825-137~\citep{aharonian06a},~PSR B0833-45/Vela X~\citep{aharonian06b} and PSR J2229+6114/VER J2227+608~\citep{acciari09}.~These objects show extended TeV emission which is both larger than the X-ray PWN and offset from the pulsar location~(see, e.g, \cite{deJager08}).~A common element in this class is older, middle-aged pulsars, with ages $>$10$^4$ yr.~Here the PWN may be displaced in some direction, possibly by an asymmetric SNR reverse shock generated from an explosion into an inhomogeneous medium~\citep{blondin01}. 
\par
Table 2 lists the pulsar properties (spin-down power, age and distance) and the nebular energetics in several compact PWNe detected in TeV.~The total power, L$_g$, radiated by G54.1+0.3 in the energy band 0.2 - 10 TeV is $2.1\times10^{34}$ erg s$^{-1}$ compared to $8.6\times10^{34}$ erg s$^{-1}$ for the Crab Nebula. This value of the gamma-ray luminosity of G54.1+0.3 is  the same order of magnitude what has been found for the other PWNe in the table.~A more relevant quantity to compare is the efficiency of TeV luminosity production compared to the pulsar power which is $\eta_{\gamma}\sim$0.17\% for G54.1+0.3.~This is similar to that of G0.9+0.1~($\approx$~0.1\%) and 10 times larger than other young systems like the Crab Nebula~($\approx$~0.02\%) and HESS G21.5-0.9~($\approx$~0.02\%).~However, all these PWNe have significantly lower levels~(ranging from 0.1 to 2.9\%) of X-ray luminosity production efficiency than the Crab Nebula ($\approx$~4\%).~Despite being worst synchrotron emitters than the Crab Nebula, G54.1+0.3 together with the other recent compact TeV PWNe, are similar and even better TeV emitters than the Crab Nebula.
\par
Another diagnostic is the strength of the synchrotron emission relative to the gamma-ray emission, L$_x$/L$_\gamma$.~The Crab Nebula has the largest ratio among all young systems by far with L$_X$/L$_\gamma$ of 220.~That of G54.1+0.3 is the lowest among the others that range between 1.4 and 30.~These numbers are consistent with~\citet{mattana09} who finds  L$_\gamma$$>$L$_X$ after 5 kyr from pulsar birth,~including a different list of objects and spanning over a larger range of $\tau_c$.~The scattering of this ratio in two orders of magnitude, may be attributed to the unique environments in each of these objects, such as the local ambient photon fields or the interactions with the surrounding mediums resulting in an increase of the nebular magnetic field.~As suggested by~\cite{chevalier04, djannati07}, these PWNe may not be in energy equipartition between particles and magnetic fields, as is the case for the Crab Nebula.~Instead, these may be particle-dominated objects.~However, proper modeling of each of these PWNe and its surroundings is needed to constrain the nebular magnetic fields.
\par
Given the association by~\citet{leahy08} of the PWN with a molecular CO cloud,~an alternative interpretation of the TeV emission from G54.1+0.3 would be radiation from the decay of neutral pions produced in the hadronic interactions produced between the dense target and the SNR. To establish if this is a compatible scenario, deeper observations of this source would be needed to constrain the size and location of the accelerated particles in the TeV range.~A detection by Fermi-LAT would add data to the spectral energy distribution of the system, giving an idea of the possible role played by protons by showing (or not) the typical bump in the spectrum expected from neutral pions (e.g.~\citet{aharonian94}).~However, based on the association with a high spin-down pulsar, a comparable size of the X-ray and radio PWNe given the instrument resolutions and comparison with similar detected objects, the PWN scenario best explains the data.~Under the assumption that the gamma-ray production in VER J1930+188 is IC emission, the recently discovered IR shell of ejecta is an additional source of seed photons from the local dust up-scatted into the TeV energies and contributing in the observed emission.  

\acknowledgements
This research is supported by grants from the U.S. Department of Energy, the U.S. National Science Foundation and the Smithsonian Institution, by NSERC in Canada, by Science Foundation Ireland and by STFC in the UK.

{\it Facilities:} \facility{VERITAS}.

\bibliographystyle{apj}

\end{document}